\documentclass[reprint,aps,prd,amssymb]{revtex4-2}

\usepackage{graphicx}
\graphicspath{{Figures/}}

\usepackage{mathtools}
\usepackage[svgnames]{xcolor} %OMG - use SVG, not DVI or ODP or PS or POOP

\usepackage[colorlinks=True, linkcolor=SlateBlue,
            citecolor=SteelBlue,urlcolor=BlueViolet]{hyperref}
\usepackage[capitalise]{cleveref}

\usepackage{aas_macros}

\newcommand{\vect}[1]{\boldsymbol{#1}}
\newcommand{\uvect}[1]{\vect{\hat{#1}}}
\newcommand{\la}{\langle}
\newcommand{\ra}{\rangle}

\defcitealias{Martynov:17}{Phys. Rev. A 95, 043831 (2017)}

\begin{document}

\author{Hang Yu}\thanks{hangyu@caltech.edu}
\affiliation{TAPIR, Walter Burke Institute for Theoretical Physics, Mailcode 350-17\\ California Institute of Technology, Pasadena, CA 91125, USA}

\author{Denis Martynov}
\affiliation{School of Physics and Astronomy, and Institute of Gravitational Wave Astronomy,\\
University of Birmingham, Edgbaston, Birmingham B15 2TT, United Kingdom}

\author{Rana X Adhikari}
\affiliation{LIGO Laboratory, California Institute of Technology, MC 100-36, Pasadena, CA 91125, USA}%

\author{Yanbei Chen}
\affiliation{TAPIR, Walter Burke Institute for Theoretical Physics, Mailcode 350-17\\ California Institute of Technology, Pasadena, CA 91125, USA }

\title{Exposing Gravitational Waves below the Quantum Shot Noise}

\begin{abstract}
% \begin{itemize}
% \item Finding signals with unknown waveforms is harder than known.
% \item We have a new way to find signals in a sort of waveform independent way.
% \item Our technique goes beyond classical signal processing by usng quantum woo woo.
% \item Using our technique, we find that we can detect BLAH BLAH BLOOP way better than anyone else, and this is really cool for some reason.
% \end{itemize}

The sensitivities of ground-based gravitational wave (GW) detectors are limited by quantum shot noise at a few hundred Hertz and above. 
Nonetheless, one can use a quantum-correlation technique proposed by Martynov, et al. [\citetalias{Martynov:17}] to remove the expectation value of the shot noise, thereby exposing underlying classical signals in the cross spectrum formed by cross-correlating the two outputs in a GW interferometer's anti-symmetric port. 
We explore here the prospects and analyze the sensitivity of using quantum correlation to detect astrophysical GW signals.
Conceptually, this technique is similar to the correlation of two different GW detectors as it utilizes the fact that a GW signal will be correlated in the two outputs but the shot noise will be uncorrelated. 
Quantum correlation also has its unique advantages as it requires only a single interferometer to make a detection. Therefore, quantum correlation could increase the duty cycle, enhance the search efficiency, and enable the detection of highly polarized signals. 
In particular, we show that quantum correlation could be especially useful for detecting post-merger remnants of binary neutron stars with both short ($< 1\,{\rm s}$) and intermediate ($\sim 10-10^4\,{\rm s}$) durations and setting upper limits on continuous emissions from unknown pulsars. 
\end{abstract}

\maketitle

\section{Introduction}
\label{s:intro}
To date, nearly 100 gravitational-wave (GW) events have been detected~\cite{GWTC1, GWTC2, GWTC3, Venumadhav:20, Olsen:22} by ground-based interferometers including Advanced LIGO (aLIGO; \cite{TheLIGOScientific:2014jea}), Advanced Virgo \cite{TheVirgo:2014hva}, and KAGRA~\cite{Akutsu:2018axf, Akutsu:21}. 

The most statistically powerful way to make a detection employs a technique known as matched filtering~\cite{Cannon:10, Allen:12, Messick:17, Venumadhav:19}. 
However, this technique has a potential limitation in that it requires accurate waveform templates. While this can be achieved for the inspiral stage of coalescing compact binaries, there are other potential GW sources whose theoretical waveforms might still have large theoretical uncertainties or be challenging to be constructed. This includes the post-merger signal of a binary neutron star (BNS) event (see, e.g., Ref.~\cite{Abbott:17, LVC:19longBNSpost, GW170817c} and references therein). Other possibilities include the GW emission from core-collapsing supernovae~\cite{LVK:21longburst, LVK:21shortburst, vanPutten:19}, accretion disk instabilities~\cite{vanPutten:01}, eccentric binary coalescence~\cite{Huerta:18, LVC:19ebbh}, 
%pulsars \Rana{cite some CW thing here}, 
% HY: pulsars' waveform are typically ``known'' at least for the CW searches considered by LIGO. Therefore they don't belong to this section sepecifically. 
etc. See also Ref.~\cite{LVK:21longburst} and references therein. 

Detection of these types of sources, therefore, calls for waveform-agnostic detection methods that do not assume a waveform template \emph{a priori}. Multiple search algorithms for unmodeled GW signals have been developed following different principles, and examples of this family of algorithms include Coherent Wave Burst~\cite{Klimenko:16}, Stochastic Transient Analysis Multidetector Pipeline~\cite{Thrane:11}, X-Pipeline~\cite{Sutton:10}, BayesWave~\cite{Cornish:15, Cornish:21},  etc.

We present here a complementary method to the family of un-modeled burst search algorithms. This method utilizes a quantum correlation technique~\cite{Martynov:17}. In the current LIGO configuration, the optical signal produced by the main interferometer is split on to two photodetectors (PDs). Intuitively, a signal field produced by physical motions in the interferometer will be correlated among the two PDs whereas the quantum shot noise will be uncorrelated. One can thus remove the quantum shot noise by cross-correlating the two PDs outputs. This is in close analogy to how a GW signal could be detected by cross-correlating two different interferometers~\cite{Allen:99, Thrane:11}, except for that the correlation now requires only a single interferometer.  

Quantum correlation has previously been used to constrain classical noise sources in LIGO~\cite{Martynov:17, Buikema:20}. In this work, we further explore the possibility of applying it to detect astrophysical GW events. As we will see later, quantum correlation can be especially beneficial for the search of signals associated with NSs as they typically reside at high frequencies ($\gtrsim 100\,{\rm Hz}$) where quantum shot noise limits a ground-based interferometer's sensitivity.

The rest of the paper is organized as follows. 
In \Cref{sec:QC}, we review the basics of quantum correlation and draw its connection with two-interferometer correlation to establish the signal-to-noise ratio for our subsequent analysis.  
This is followed by \Cref{sec:QC_vs_2IFO} where we further discuss the potential benefits of quantum correlation. 
Then in \Cref{sec:astro} we consider applying quantum correlation to detect astrophysical signals, including post-merger remnants of BNS events with short ($<1\,{\rm s}$; \Cref{sec:short_burst}) and intermediate ($\sim 100\,{\rm s}$; \Cref{sec:long_burst}) durations and continuous-wave emissions from Galactic pulsars (\Cref{sec:cw}). 
Lastly, we conclude our study and discuss its limitations and future directions in \Cref{sec:conclusion_discussion}.

\section{Removing quantum shot noise}
\label{sec:QC}

\begin{figure}
  \centering
  \includegraphics[width=\columnwidth]{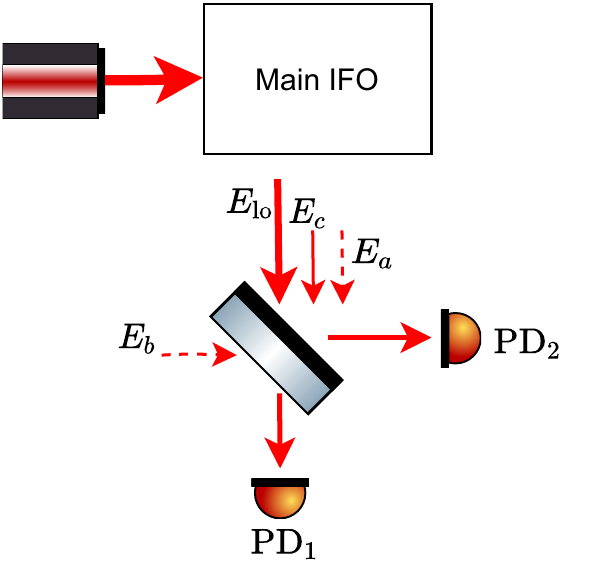}
  \caption{Cartoon illustrating the readout port of an aLIGO-like GW detector.   }
\label{fig:cartoon}
\end{figure}

We begin our discussion by reviewing how one may remove the quantum shot noise in the readout spectrum using the cross-correlation technique described in \cite{Martynov:17}. We will assume first that there is no quantum squeezing injected into the interferometer, which represents the aLIGO configuration in the first and second observing runs.
%, as well as in the third observing run for the LIGO Hanford observatory. 
We will later discuss how one may extend this method to cope with squeezed light~\cite{Tse:19}. In this work, we use the convention that we use $E$ to denote an optical field and $n$ the power response of PD. We choose the physical constants such that $\left[E^\ast E\right]$ has the unit of $\left[{\rm W}\right]$. 

For aLIGO, the signal field leaving the interferometer's anti-symmetric port is split on to two different PDs as shown in \Cref{fig:cartoon}. 
In a semiclassical way, we can write the power fluctuation on each PD as 
\begin{equation}
    n_{1,2} = {\rm Re}\left[E_{\rm lo}^\ast \left(E_a \pm E_b + E_c\right)\right],
\end{equation}
where the fields are defined as in \Cref{fig:cartoon}. Specifically, $E_{\rm lo}$ is a local-oscillator field produced by an intentional detuning of the differential arm length (which is known as the ``DC readout scheme''~\cite{Fricke:12}). 
%The fields $E_a$ and $E_b$ are vacuum fluctuations returning from the main interferometer and entering at the final beamsplitter, respectively. 
The field $E_a$ corresponds to vacuum fluctuations entering the interferometer from the anti-symmetric port and then returning to the readout PDs. Its interference with $E_{\rm lo}$ produces the quantum shot noise limiting a ground-based detector's high-frequency sensitivity. 
As we split the signal onto two different PDs, another vacuum field $E_b$ enters from the open port of the beamsplitter and it carries the same amount of fluctuation as the $E_a$ field when there is no squeezing. 
Lastly, the field $E_c$ is produced by classical differential arm length changes. At frequency $\gtrsim 100\,{\rm Hz}$, classical noises (e.g., coating thermal noise and gas phase noise) are expected to be small, and therefore $|E_c|$ is significant only when a high-frequency GW signal is present. The transfer function from a GW strain signal $h$ at frequency $f$ to the power fluctuation is given by the interferometer's optical response. It reads~\cite{Martynov:17}
%Classical differential arm motion (including a GW signal $h$) produces the field $E_c$ via the interferometer's optical response (under the  long-wavelength approximation)
\begin{align}
    Z(f) & \equiv  2 |E_{\rm lo}| \frac{dE_c}{dL}(f) \frac{dL}{dh}(f), \nonumber \\
    &=|E_{\rm lo}| \frac{4\pi}{\lambda}
    \left(G_{\rm a}G_{\rm p}P_{\rm in}\right)^{1/2}
    \frac{\tilde{t}_{\rm i}}{1-\tilde{r}_{\rm i}e^{-i4\pi f L/c }} \frac{dL}{dh}(f),
    \label{eq:dPdh}
    % &= |E_{\rm lo}| \frac{4\pi G_{\rm arm}G_{\rm prc}^{1/2}P_{\rm in}^{1/2}}{\lambda G_{\rm src}^{1/2}}\frac{f_-}{f_- + i f} \frac{dL}{dh}(f), 
\end{align}
% where the notation follows Ref~\cite{Martynov:17}. 
where $L$, $\lambda$, and $P_{\rm in}$ are, respectively, the arm length, the laser wavelength, and the input power. The factor of 2 before $E_c$ is because the GW signal is readout from the SUM channel $n_{\rm sum}=n_1 + n_2$. The factors $G_{\rm a}$ and $G_{\rm p}$ are the power buildup factors in the arm and power-recycling cavities, respectively. We further define
\begin{equation}
    \tilde{r}_{\rm i} = \frac{r_{\rm i} - r_{\rm s}}{1-r_{\rm i}r_{\rm s}},
    \text{ and }
    \tilde{t}_{\rm i} = \frac{t_{\rm i} t_{\rm s}}{1-r_{\rm i}r_{\rm s}},
\end{equation}
where $t_{\rm i}$ and $t_{\rm s}$ are the amplitude transmissivities of the input test mass and the signal-recycling mirror, and the reflectivities are given by $r_{\rm i(s)}^2\simeq 1-t_{\rm i(s)}^2$ when the optical losses are small. 
For aLIGO with an arm length of $L\simeq4\,{\rm km}$, we further have $dL/dh \simeq L$ and $\exp\left(-i4\pi f L/c \right)\simeq 1-i4\pi f L/c$. 
This leads to 
\begin{equation}
    Z(f)\simeq |E_{\rm lo}| \frac{4\pi}{\lambda} \frac{G_{\rm arm}G_{\rm prc}^{1/2}P_{\rm in}^{1/2}}{G_{\rm src}^{1/2}}\frac{f_-}{f_- + i f} L, 
    \label{eq:dPdh_long_wave}
\end{equation}
where $f_-\simeq \left(1-\tilde{r}_i\right) c/(4\pi L)$ is the coupled-cavity pole frequency and $G_{\rm src}\simeq \left(1+r_s\right)^2/t_s^2$. Note that Eq.~(\ref{eq:dPdh_long_wave}) reduces to the expression derived in Ref.~\cite{Martynov:17}. For the future generation of GW detectors like the Cosmic Explorer (CE; \cite{Evans:17, Evans:19, Evans:21}) with $L\simeq 40\,{\rm km}$, the approximation breaks down and therefore the full expression, Eq.~(\ref{eq:dPdh}), is used with the $dL/dh(f)$ term given by Ref.~\cite{Schilling:97} (as done in noise budgeting codes like \texttt{pygwinc}~\footnote{\url{https://git.ligo.org/gwinc/pygwinc}}).

If we denote the power spectral density (PSD) of ${\rm Re}\left[E_{\rm lo}^\ast E_{a,b}\right]$ as $P_{a,b}$ (whose expectation is $\la P_a \ra = \la P_b \ra =|E_{\rm lo}|^2 h\nu/2$ and units are $[{\rm W^2\,Hz^{-1}}]$), then the PSD of the SUM channel  due to the quantum shot noise (i.e., without $E_c$) is $P_{\rm sum} = 4 P_a$, and the PSD of the shot noise in terms of the GW strain can be obtained by $\la P_{\rm sum} \ra /|Z(f)|^2$. Similarly, we can define a NULL channel as $n_{\rm null}=n_1-n_2$ and its PSD is $P_{\rm null}=4P_b$.  

As shown in Ref.~\cite{Martynov:17}, we can get rid of the quantum shot noise through a quantum-correlation technique. Specifically, we compute the real part of the cross spectral density (CSD) of $n_1$ and $n_2$, which we denote as $C_{12}\equiv{\rm Re}\left[{\rm CSD}\left(n_1, n_2\right)\right]$, and its expectation is given by
\begin{equation}
    \la C_{12} \ra = \la P_a \ra - \la P_b \ra +  \la P_c\ra = \la P_c\ra.
    \label{eq:C12_exp}
\end{equation}
We thus see the shot noise is removed in the CSD and we are left with only the classical length changes of the interferometer, $\langle P_c\rangle$. On the other hand, we note the cancellation is done in the expectation. For a specific realization of $C_{12}$ (i.e., a pixel in the spectrogram or the $ft-$map), the variance is given by 
\begin{equation}
    \sigma_{C_{12}}^2 =\frac{1}{2}\left(\la P_a \ra +  \la P_b \ra\right)^2 = 2\la P_a \ra^2=2\la P_b \ra^2.
    \label{eq:sigma_csd_pixel}
\end{equation}
The signal-to-noise ratio (SNR) of each pixel is thus
\begin{equation}
    \rho = \frac{C_{12}}{\sigma_{C_{12}}} \text{, and } 
    \la \rho \ra = \frac{\la P_c \ra}{\sqrt{2} \la P_a\ra } \propto \frac{h^2}{\la P_a \ra/|Z|^2 }. 
    \label{eq:snr_per_pixel}
\end{equation}
This is similar to how the SNR is defined in the case of two-interferometer correlation~\cite{Thrane:11}. 
Note that the SNR is inversely proportional to the PSD of the shot noise $\la P_a \ra/|Z|^2$ (which has a unit of $\left[{\rm strain^2/Hz}\right]$). 

In fact, one can also remove the shot noise in expectation by computing the PSD of the NULL channel and then subtract it from the PSD of the SUM channel, 
\begin{equation}
    \la P_{\rm diff}\ra \equiv \la P_{\rm sum} \ra - \la P_{\rm null} \ra = 4\left(\la P_a \ra - \la P_b \ra +  \la P_c\ra\right) = 4\la P_c\ra.
\end{equation}
The variance on each pixel is 
\begin{equation}
    \sigma_{\rm diff}^2 = \sigma_{\rm sum}^2 + \sigma_{\rm null}^2 =32 \la P_a \ra^2.
\end{equation}
We immediately see that the SNR obtained this way will be the same as the one obtained from the correlation technique. 

How does the SNR of the quantum correlation technique compare with the one obtained by the cross-correlating two different interferometers (see, e.g., \cite{Thrane:11})? In the two-interferometer correlation scenario, we note that the signal's contribution to the CSD is $4 P_c$ if we assume the two interferometers have identical configurations and signal responses. The standard deviation due to uncorrelated detector noise sources is $2\sqrt{2} \la P_a \ra$ in the CSD (assuming in the shot noise limited regime). Therefore the SNR is $\sqrt{2}  P_c  / \la P_a \ra$ for two-interferometer correlation. 
On the other hand, if we perform quantum correlation on each interferometer individually first and then combine the SNR [Eq.~(\ref{eq:snr_per_pixel})] in quadrature, the SNR is $P_c / \la P_a \ra$, which is $\sqrt{2}$ lower than directly correlating the two interferometers' outputs. The physical reason for this degradation is the following. As we utilize the  $E_b$ vacuum field to cancel out the expectation of the $E_a$ field that causes the shot noise in the SUM channel, we inevitably introduce the fluctuations associated with $E_b$ to the system as well. Therefore, the SNR is degraded and this is also the reason why a $\sqrt{2}$ appeared in the denominator of \eqref{eq:snr_per_pixel}. Despite of the loss in the SNR, the quantum correlation technique nonetheless has a few unique advantages thanks to the fact that the signal field is produced by a single interferometer, which we will discuss in more details in \Cref{sec:QC_vs_2IFO}. Therefore, it is still interesting to consider its application in detecting astrophysical signals in \Cref{sec:astro}.

% \hang{
% HY: if the LO field does not have much variations above 1 Hz, then it seems we can just use the mean $E_{\rm LO}$ every second, compute $4\la P_a \ra$, and subtract it from the SUM channel directly? If $E_{\rm LO}$ is known perfectly, then we are just left with $\sigma_{\rm sum} = P_{\rm sum} =4 \la P_a \ra$, and the SNR in this case will be $\la P_c \ra / \la P_a \ra$ which is $\sqrt{2}$ greater than Eq.~(\ref{eq:snr_per_pixel}). I.e., we are no more subject to the fluctuation in $E_b$.}

% \hang{
% If we know the source's sky location,  it seems we can further cross-correlate PD 1 from LHO with PD 1 from LLO, etc., as long as we do proper corrections on the antenna response.  In this case we have 6 pairs in total, and we can thus increase the SNR in Eq.~(\ref{eq:snr_per_pixel}) by $\sqrt{6}$. This will then exceed the sensitivity by cross-correlating two SUM channels by a factor of $\sqrt{3/2}$. Is this too good to be true?
% }

When $E_a$ is squeezed and $E_b$ is not, we see that the shot noise does not vanish in the CSD as shown in Eq.~(\ref{eq:C12_exp}). % as $\la P_a \ra \neq \la P_b \ra$. 
Nonetheless, we can remedy the situation by also squeezing the $E_b$ field such that we again have $\la P_a \ra = \la P_b \ra$.  
Note that this condition is needed only in the band where the shot noise dominates. Therefore, while the $E_a$ field is anticipated to be squeezed in a frequency-dependent way (e.g., via a filter cavity~\cite{McCuller:20, Zhao:20}), a frequency-independent squeezing source is sufficient for the $E_b$ field to achieve $\la P_a \ra = \la P_b \ra$ at $f\gtrsim \text{a few }\times100\,{\rm Hz}$. 
Therefore, quantum correlation may be used not only for the archival LIGO data where the analysis above readily applies (as it has been used to constrain classical noise sources~\cite{Martynov:17, Buikema:20}), but also for future detectors like LIGO-A+~\cite{LVK:18obsscen}, LIGO-Voyager~\cite{Adhikari:17, Adhikari:20}, LIGO-HF~\cite{Martynov:19}, the Neutron Star Extreme Matter Observatory~\cite{Ackley:20}, the Einstein Telescope~\cite{Hild:10, Punturo:10, Hild:11, Sathyaprakash:12}, and CE~\cite{Evans:17, Evans:19, Evans:21} if an additional squeezed vacuum source would be installed for $E_b$ field so that $\la P_a \ra = \la P_b \ra$. We will assume this to be the case when applying quantum correlation for future detectors. 
%We thus also consider applying the quantum correlation technique for future detectors like CE under the assumption that the $E_b$ field will be squeezed by the same amount as the $E_a$ field.

% When we analyze the sensitivity of this scheme for future detectors such as CE, we assume $E_b$ will be squeezed by the same amount of $E_a$. 

\section{Comparison with two-interferometer correlation}
\label{sec:QC_vs_2IFO}

Conceptually, we note quantum correlation is largely similar to the technique of cross-correlating two different interferometers~\cite{Thrane:11}. They both utilize the fact that the signal is correlated among different readout channels whereas the noise is uncorrelated. Compared to two-interferometer correlation, quantum correlation has the drawback that it only removes the quantum shot noise. And even in the shot-noise limited band, its sensitivity is slightly degraded due to the introduction of a new vacuum field into the system [which leads to the $\sqrt{2}$ factor in the denominator in Eq.~(\ref{eq:snr_per_pixel})]. Nonetheless, quantum correlation has a few unique advantages thanks to the fact that \emph{it requires only a single interferometer}. 

First of all, a single detector naturally means a higher duty cycle compared to coincident observation between at least a pair of interferometers as required by the two-interferometer correlation. For instance, during the third observing run, each LIGO detector achieved a duty cycle of $\simeq 75\%$ individually, and the joint observation covered $\simeq 60\%$ of the time~\cite{Buikema:20}. A higher duty cycle means that it is less likely for us to miss an astrophysical signal especially if the signal is transient in nature. 
Quantum correlation can thus be a critical backup plan for methods originally requiring two interforemeters in case that only one detector is online during a GW event. 

Quantum correlation could also make searches for GW events more efficient, as to detect a signal from a single detector, one does not need to know the source's sky location. 

To see this advantage, we first briefly review the basics to perform two-interferometer correlations~\cite{Allen:99, Thrane:11}. 
Note that the GW strain observed by an interferometer $I$ can be written as 
\begin{equation}
    h_I(t + \tau_I) = F_{I+}(\uvect{\Omega}, \psi)h_+(t) + F_{I\times}(\uvect{\Omega}, \psi)h_\times(t)
\end{equation}
where $h_{+(\times)}$ is the waveform in the $+(\times)$ polarization,  $F_{I+(\times)}$ is the antenna response of interferometer $I$ to each polarization and it depends on the sky location of the source $\uvect{\Omega}$ and the polarization angle $\psi$. We further use $t$ to denote the time when the GW wavefront arriving at a reference point and $\tau_I(\uvect{\Omega})$ the time for the wave to propagate from the reference to detector $I$. 
To perform two-interferometer correlation, one would need to account for the difference in a signal's arrival time and antenna responses in interferometers $I$ and $J$, by applying a correction factor to align their outputs~\cite{Thrane:11}
\begin{equation}
    % Q_{IJ} = \frac{2}{F_{I+}F_{J+} + F_{I\times}F_{J\times}}e^{2\pi i f \uvect{\Omega}\cdot \vect{r}_{IJ}/c}.
    Q_{IJ} = \frac{2}{F_{I+}F_{J+} + F_{I\times}F_{J\times}}e^{2\pi i f \Delta \tau_{IJ}},
\end{equation}
where $\Delta \tau_{IJ}$ is the time delay of the signal in different detectors. It can be further evaluated as $\Delta \tau_{IJ}= \uvect{\Omega}\cdot \vect{r}_{IJ}/c$, where $\vect{r}_{IJ}$ is the difference in position vectors of detectors $I$ and $J$.
For a source either with unknown location or poorly localized, we need to search over a large portion of the sky in order to perform cross-correlation between two interferometers. This could be a task computationally expensive.

For quantum correlation, this correction is not needed as the signal will reach the two readout PDs at effectively the same time (as $\Delta \tau_{II}=0$) and the antenna response can be absorbed into an effective distance~\cite{Allen:12}. As a result, we would only need to search over the intrinsic source parameters, thereby reducing the computational cost. 
Note, however, that this does not mean we discard the information on the source's sky location. If we can detect the signal in two different interferometers, we can then readout the time delay between the interferometers as well as the difference in the SNR to infer the source's sky position. 
In other words, the sky location is inferred \emph{after} the GW event's detection. 
Alternatively, we can also use quantum correlation as a first step to achieve the detection and to constrain the intrinsic source parameters, and then follow it up with more sensitive yet more computationally expensive analyses.

% 3. Works for signals whose $h_+$ and $h_\times$ are not highly correlated. For general astrophysical bursts without much symmetry, this may be the case. If one IFO is sensitive only to $h_+$ and the other only to $h_\times$, then the cross-correlation of the two simply won't give a signal. 

Yet another advantage of quantum correlation is that it works for polarized signals. Imagine an extreme example. If interferormeter $I$ is sensitive only to the $+$-polarization and $J$ only to the $\times$-polarization ($F_{I+} {=} F_{J\times} {=} 1$,  $F_{I\times} {=} F_{J+} {=} 0$), then the cross-correlation between the two would not be able to detect a polarized signal with only the $h_+$ component. Nonetheless, with quantum correlation, we would at least be able to detect the signal in interferometer $I$. 

Given these advantages, we investigate how quantum correlation may help us detect astrophysical GW signals in the following section. 

\section{Astrophysical applications}
\label{sec:astro}

%We consider here the application of quantum correlation to various astrophysical sources. 
Quantum correlation enables removing the shot noise but not classical noise, it is mostly valuable for the detection of various GW signals related to NSs at shot-noise limited frequencies, or $f\gtrsim100\,{\rm Hz}$ for ground-based GW observatories. In particular, quantum correlation could potentially help the detection of post-merger bursts of BNS events with short ($< 1\,{\rm s}$) or intermediate $\left[\sim \mathcal{O}(100)\,{\rm s}\right]$ duration (Secs.~\ref{sec:short_burst} and \ref{sec:long_burst}), as well as persistent or continuous GW emission from unknown pulsars (\Cref{sec:cw}).  

As our focus is to detect the signal and extract the intrinsic parameters using a single detector, we will fix the source to have a face-on orientation and an antenna response $F_+=1,\ F_\times=0$ for the cases in Secs.~\ref{sec:short_burst} and \ref{sec:long_burst}. Changing the source's orientation, sky location, and polarization only affects the overall amplitude of the signal. Note, however, that these extrinsic parameters (such as the source's sky location) can be extract \emph{after} the detection has been made in \emph{multiple} detectors, though we deferred the analysis on extrinsic parameters to future studies.

\subsection{Short-duration bursts}
\label{sec:short_burst}

\begin{figure*}
  \centering
  \includegraphics[width=1.95\columnwidth]{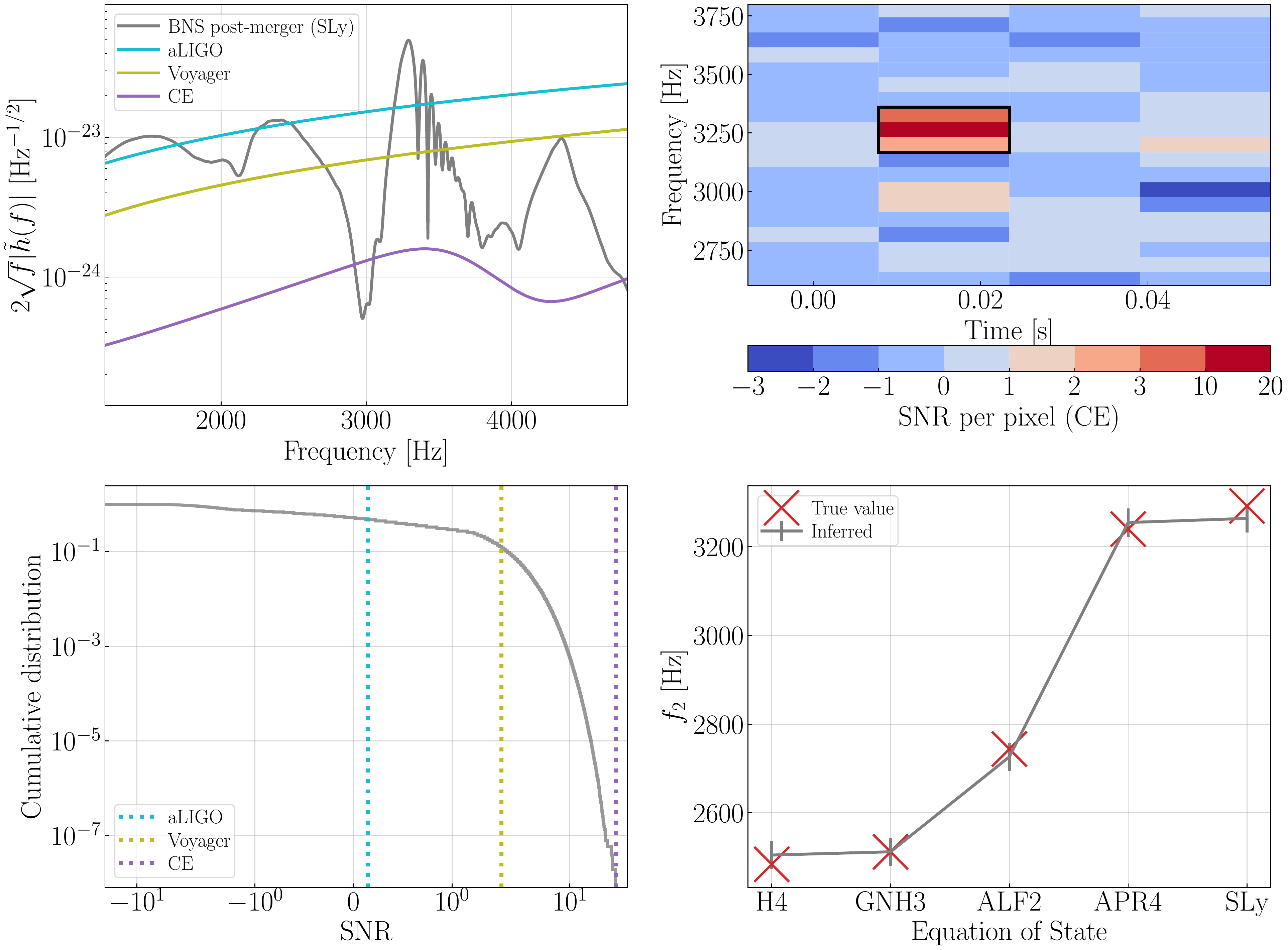}
  \caption{
  Top-left: comparison between the characteristic strain of a BNS post merger signal (gray curve) and the sensitivity of representative GW detectors (colored curves). The source is placed at a distance of $100\,{\rm Mpc}$ with a face-on orientation and $F_+=1,\ F_\times=0$.
  Top-right: example of an $ft$-map formed by quantum correlation containing the post-merger signal (the pre-merger part is filtered out). The detector is assumed to have CE's sensitivity. To detect the signal, we search over boxes spanning $(1/64\,{\rm s}, 192\,{\rm Hz})$ (i.e., three pixels), and the box leading to the highest total SNR is highlighted by with a black boundary. 
  Bottom-left: cumulative distribution of background (signal-free) boxes with SNR greater than a value given by the x-axis. The vertical lines are the expected SNRs of the signal in the top-left panel in different detectors under the quantum correlation technique. 
  Bottom-right: Comparison between the inferred $f_2$ frequency (black-pluses) and the true $f_2$ frequencies for different NS equation of states. 
  %Top: ft-map containing a short BNS post-merger signal. We compute the SNR of each pixel according to Eq.~(\ref{eq:snr_per_pixel}) and search for the post-merger over boxes like the the one whose boundaries highlighted by the thick black lines.  Bottom-left: Spectrum of the signal (gray) and the sensitivity of typical GW detectors. Bottom-right: Cumulative distribution of the mean SNR over each box of the background (gray). Shown in the vertical lines are the expected SNRs in GW detectors.  \Rana{ADD VOYAGER!}
  }
\label{fig:short_burst}
\end{figure*}

% BNS post-merger signal with duration less than 1 s (typically only $\sim \mathcal{O}(10)\,{\rm ms}$).
It has been postulated that if a BNS's total mass exceeds a maximum value allowed by a uniformly rotating star, its post-merger remnant could be a hypermassive NS that collapse in less than a second~\cite{Baumgarte:00, Abbott:17}. GW170817~\cite{GW170817}, for example, is likely to have a total mass that can lead to a hypermassive NS~\cite{Abbott:17, GW170817b, GW170817c}. 

The associated GW waveform of a hypermassive NS has been extensively studied by the literature (see, e.g., Refs.~\cite{Hotokezaka:13, Takami:14, Takami:15, Rezzolla:16,Kawamura:16, Dietrich:18, Zappa:18}). For the analysis here, we take GW waveforms provide in \cite{Takami:14, Takami:15, Rezzolla:16}. 
The comparison between the GW signal (the gray curve) and the shot noise level of representative detectors (colored curves) are shown in the top-left panel in Fig.~\ref{fig:short_burst}. We use cyan, olive, and purple to respectively represent aLIGO, Voyager, and CE and this convention will be used throughout this Section. 
We place the merger at a distance of $100\,{\rm Mpc}$ and assume the SLy equation of state~\cite{Douchin:01}. 
The waveform is filtered the same way as described in Ref.~\cite{Takami:15} to remove the pre-merger part.

To perform the quantum correlation measurement, we convert the strain signal to power fluctuations on the PDs using Eq.~(\ref{eq:dPdh}) and superpose it with the other noise sources (dominated by the quantum shot noise in the band of interest). We then normalize each pixel by the standard deviation $\sigma_{C_{12}}$ [Eq.~\ref{eq:sigma_csd_pixel}]. A resultant normalized $ft$-map is presented in the top-right panel of \Cref{fig:short_burst}. In this example, the merger happens at $t=0.018\,{\rm s}$.

To detect the signal, we consider detection boxes with a size of $(1/64\,{\rm s}, 192\,{\rm Hz})$, corresponding to three pixels in the $ft$-map (see, e.g., the black box in the top-right panel of \Cref{fig:short_burst}). The cumulative distribution of the total SNR in many realizations of signal-free boxes is further shown in the lower-left panel of \Cref{fig:short_burst}, serving as the background statistics for us to construct the detection threshold. 
Simulations here are performed over simulated Gaussian noises. 
Also shown in the vertical lines are the expected SNR inside the detection box of the signal shown in the top-left panel in various detectors.

Following Ref.~\cite{Abbott:17}, we define the root-sum-squared strain amplitude as
\begin{equation}
    h_{\rm rss} = \sqrt{2\int \left(|\tilde{h}_+(f)|^2 + |\tilde{h}_\times(f)|^2\right) df},
\end{equation}
where $\tilde{h}_{+(\times)}(f)$ are the Fourier transforms of $h_{+(\times)}(t)$. The efficiency of the algorithm is then analyzed in terms of the minimum $h_{\rm rss}$ required in order to make the false-alarm probability (FAP) lower than a certain threshold. Because our detection box is about 100 times shorter than the one used in Ref.~\cite{Abbott:17}, we thus choose a threshold of ${\rm FAP}=10^{-6}$ which is 100 times lower than the threshold used in Ref.~\cite{Abbott:17}. This leads to $h_{\rm rss}=1.5\times10^{-22}$, $3.8\times10^{-23}$, and $1.1\times10^{-23}$ for the short-duration post-merger signal to be detected by aLIGO, Voyager, and CE, respectively. If we instead choose a more strict detection threshold of ${\rm FAP}=10^{-7}$, this only increases the value of $h_{\rm rss}$ by $6\%-7\%$. 

Besides detecting the signal itself, it is also of great significance to constrain the peak frequency of the post-merger signal. This is also known as the $f_2$ frequency following the convention used in Ref.~\cite{Rezzolla:16} and it has been shown to contain critical information of the NS equation of state~\cite{Shibata:05}. The $f_2$ peak frequency can be constrained from the $ft$-map by computing the SNR-weighted-mean frequency of the detection box that has the maximum SNR. 

In the bottom-right panel of Fig.~\ref{fig:short_burst}, we compare the inferred $f_2$ frequency (gray pluses) and the true $f_2$ frequency for a variety of NS equation of states, covering hard (H4~\cite{Glendenning:91}; GNH3~\cite{Glendenning:85}), intermediate (ALF2~\cite{Alford:05}), and soft (APR4~\cite{Akmal:98}; SLy~\cite{Douchin:01}) ones. In all the cases, we keep the source at 100\,Mpc and assume CE's sensitivity in the detector. We find the fluctuation in the inferred value due to different noise realization is small $<10\,{\rm Hz}$ and the uncertainty in the inference is set by the resolution of the pixel. For all the equation of states we consider, the difference between the inferred value and the true one is less than half of the pixel's size, or $32\,{\rm Hz}$, which we adopt as the size of the error bar when generating the gray markers in the plot. It is thus possible for us to use quantum correlation and future detectors to distinguish hard, intermediate, and soft equation of states.

\subsection{Intermediate-duration bursts}
\label{sec:long_burst}

\begin{figure*}
  \centering
  \includegraphics[width=1.85\columnwidth]{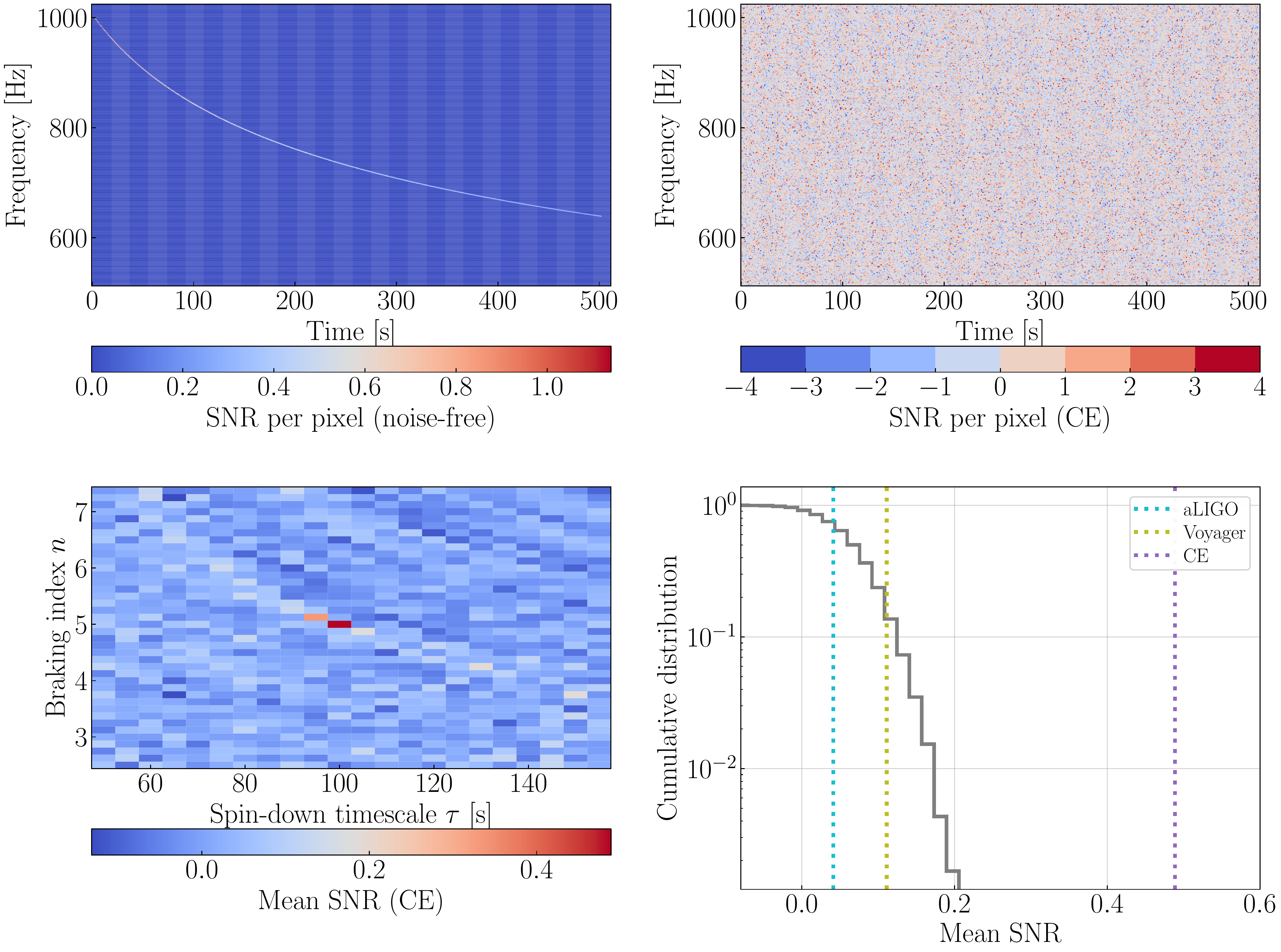}
  \caption{Top-left: $ft$-map of the spin-down signal of a magnetar without noise. Each pixel in the $ft$-map spans ($1\,{\rm s}$, $1\,{\rm Hz}$). We place the magnetar at a distance of $40\,{\rm Mpc}$ and assume it has an ellipticity $\epsilon = 10^{-2}$.  
  Top-right: $ft$-map including both the signal and background noise (assuming CE sensitivity). 
  Bottom-left: mean SNR along trajectories specified by different spin-down timescale $\tau$ and braking index $n$ [Eq.~(\ref{eq:magnetar_f_vs_t})]. When the parameters agree with the values describing the signal ($\tau=100\,{\rm s}$; $n=5$), we note a significant increase in the mean SNR, thereby allowing the signal to be detected and the intrinsic parameters to be constrained. 
  Bottom-right: cumulative distribution of background trajectories. The vertical lines are the expected SNRs of the signal in the top-left panel in different detectors under the quantum correlation technique. 
  }
\label{fig:long_burst}
\end{figure*}

% Less massive BNS mergers may lead to a long-lived post-merger product that lasts $10-10^4\,{\rm s}$. 
For less massive BNS mergers, the remnant could be a supermassive NS whose mass is greater than the maximum value for a non-rotating NS. In this case, the GW signal could have a long duration, ranging from $10\,{\rm s}$ to $10^4\,{\rm s}$~\cite{Ravi:14}.

Following Refs.~\cite{Abbott:17, LVC:19longBNSpost}, here we consider the possibility that the merger produces a fast-spinning magnetar~\cite{Cutler:02}. The subsequent spinning down of the magnetar follows a trajectory~\cite{Lasky:17}
\begin{equation}
    f(t)=f_0\left(1+\frac{t}{\tau}\right)^{1/(1-n)},
    \label{eq:magnetar_f_vs_t}
\end{equation}
where $f_0$ is an initial GW frequency, $\tau$ is the spin-down timescale, and $n$ is the braking index. 
The phase of the GW waveform is then given by
\begin{equation}
    \Phi(t) = \Phi_0 + 2\pi \int_0^{t} f(t') dt',
\end{equation}
and the overall amplitude~\cite{Lasky:17} 
\begin{equation}
    h_0(t) = \frac{4\pi^2 G}{c^4} \frac{\epsilon I}{d} f^2(t),
    \label{eq:h0_vs_ellip}
\end{equation}
where $\epsilon$, $I$, and $d$ are respectively the ellipticity, moment of inertia of the NS, and the distance to the source. 

Consistent with Ref.~\cite{Abbott:17}, we adopt $(f_0, \tau, n)=(1000\,{\rm Hz},\ 100\,{\rm s},\ 5)$ to describe the spin-down trajectory. To set the amplitude, we further use $\epsilon=0.01$, $I=150\,M_\odot\,{\rm km}^2\simeq3.0\times10^{38}\,{\rm kg\,m^2}$, and $d=40\,{\rm Mpc}$. 

The signal (normalized by $\sigma_{C_{12}}$) is shown in the top-left panel of Fig.~\ref{fig:long_burst} and its superposition with detector noise (assuming CE's sensitivity) is shown in the top-right panel. Here the $ft$-map has a pixel size of $(1\,{\rm s}, 1\,{\rm Hz})$. 

While the signal is too weak to be directly visible by eyes, it can nonetheless be detected if we search for excess power along certain trajectories including hundreds of pixels. Generic clustering and pattern-recognition algorithms (e.g., Refs.~\cite{Thrane:13, Thrane:15, Sun:19, Banagiri:19}) can be applied, yet as a proof-of-concept study, here we simply search over trajectories specified by Eq.~(\ref{eq:magnetar_f_vs_t}) but with varying parameters. Each trajectory we search spans 500 pixels. 

The resultant mean SNR along different trajectories is shown in the bottom left panel of \Cref{fig:long_burst}. When we hit the right parameters describing the signal, we note a sudden increase in the mean SNR along the trajectory. This allow us to simultaneously detect the signal and constrain its properties. 

To quantitatively establish the sensitivity, we consider the cumulative distribution of the mean SNR along signal-free trajectories. The vertical lines represent the expected SNR of the signal (top-left panel of \Cref{fig:long_burst}) in different detectors. 
If we choose an ${\rm FAP}=0.01$ as the detection threshold (which is consistent with Ref.~\cite{Abbott:17}), this leads to $h_{\rm rss}=9.6\times10^{-23}$, $2.5\times10^{-23}$, and $6.3\times10^{-24}$ for aLIGO, Voyager, and CE, respectively.

\subsection{Continuous-wave sources}
\label{sec:cw}

\begin{figure}
  \centering
  \includegraphics[width=\columnwidth]{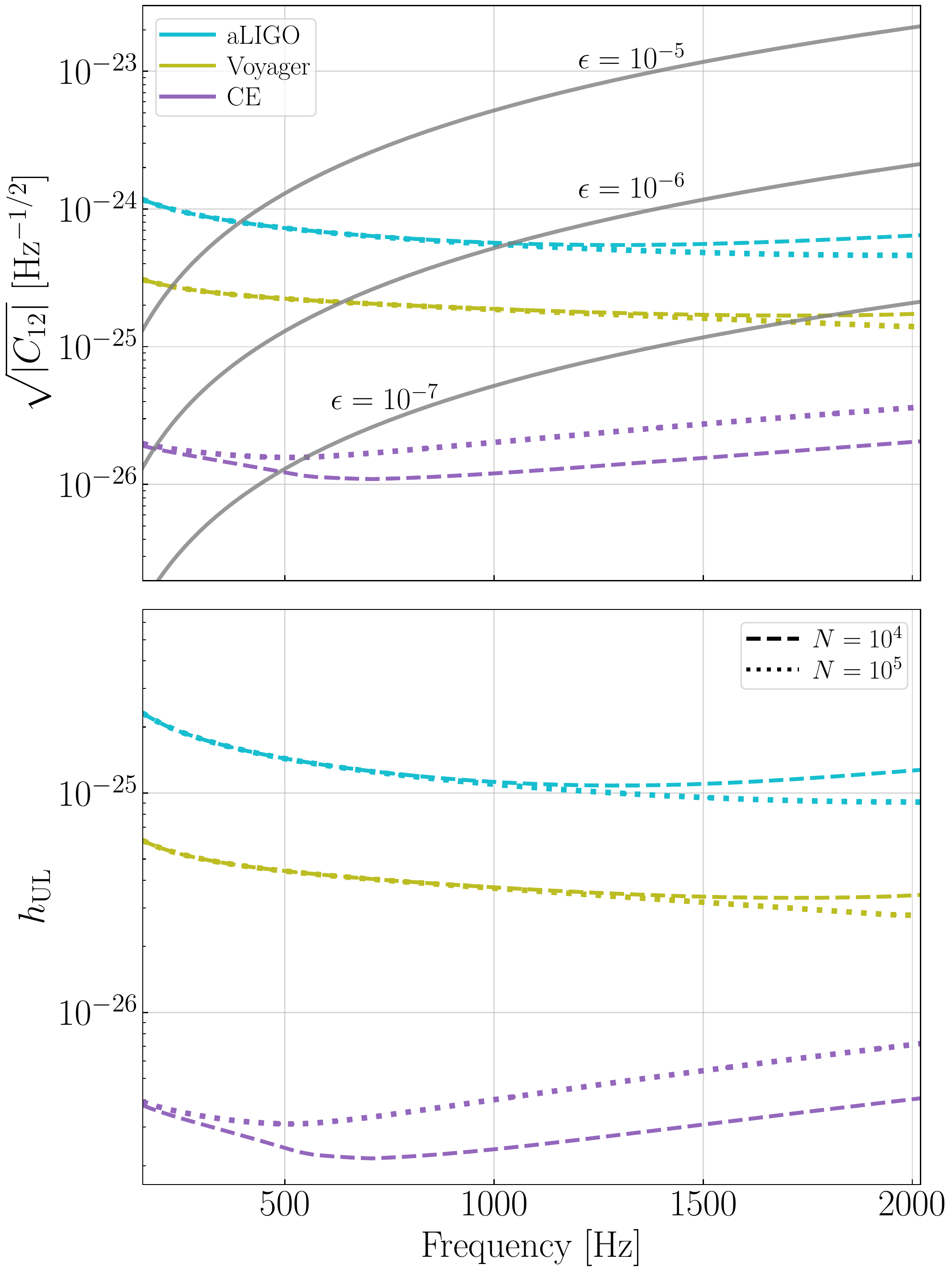}
  \caption{Top: in the colored curves, we show the expected fluctuations in the CSD after averaging $10^{4}$ (dashed) or $10^{5}$ (dotted) segments of data, respectively corresponding to total observation time of 4 months and 3 years if each segment is 1024\,s long. Different colors correspond to different detectors. Also shown in the solid traces are expected signal strength due to pulsars whose GW frequencies correspond to the x-axis. We vary the value of ellipticity as labeled in the figure, while holding the pulsar's a distance at $10\,{\rm kpc}$. Note here we have calibrated the CSD back to the strain unit. 
  Bottom: upper limit on the strain $h_{\rm UL}$ based on the expected noise level [Eq.~(\ref{eq:h_UL_CW})]. The calculation assumes $T_{\rm FFT}=1024\,{\rm s}$ and a fiducial detection threshold of ${\rm SNR_{th}}=5$. 
%   \hang{FIXME: add thermal noise \& $\sim 10\%$ mismatch in $\la P_a \ra$ and $\la P_b \ra$ due to BS imbalance or mismatches in the squeezing level, etc.}
  }
\label{fig:CW}
\end{figure}

Besides bursts, quantum correlation could potentially contribute to the search for GW emission from fast-rotating pulsars (see, e.g., ~\cite{LVC:21binpulsar,LVK:21isopulsar, LVK:22isopulsar}).

Suppose the total duration of observation is $T_{\rm obs}$. We divide the data into $N$ non-overlapping segments, and each segment has a duration of $T_{\rm FFT}=T_{\rm obs}/N$. We perform fast Fourier transfer on each segment of data and compute the corresponding CSD for quantum correlation. The expected noise in the averaged CSD over $N$ segments is approximated by the quadratic sum of the variance in the shot noise [Eq.~(\ref{eq:sigma_csd_pixel})] and the nonvanishing classical noise $P_{c, {\rm cl}}$, 
\begin{equation}
    \left(C_{12}^2\right)_{\rm noise} \approx \sigma_{C_{12}}^2/N + P_{c, {\rm cl}}^2.
    \label{eq:C12_noise_CW}
\end{equation}

The resultant noise level is shown in the top panel in \Cref{fig:CW}. We use dashed (dotted) lines to represent the estimated noise level after averaging over $N=10^{4}$ ($N=10^5$) segments. %For $T_{\rm FFT}=1024\,{\rm s}$, this corresponds to $T_{\rm obs}\simeq 1$ month ($T_{\rm obs}\simeq 3$ years) of data. 
For aLIGO and Voyager, about $10^{4}$ averages will be sufficient to reach the noise floor set by classical noise sources such as thermal noise \footnote{For simplicity, we consider here only the broadband classical noise and ignore sharp lines in the spectra.} that cannot be removed by quantum correlation here. This corresponds to $T_{\rm obs}\simeq 4\,{\rm months}$ if $T_{\rm FFT}=1024\,{\rm s}$. For CE, the classical-noise floor is significantly lower, and even with $10^5$ averages ($T_{\rm obs}\simeq 3$ years for $T_{\rm FFT}=1024\,{\rm s}$), the sensitivity is still limited by the $\sigma_{C_{12}}^2/N$ term at 1000\,Hz. 

The noise level is to be compared with the expected signal strength in the CSD which we show in the solid lines for different values of ellipticities. The amplitude of the wave can still be computed from Eq.~(\ref{eq:h0_vs_ellip}). We fix $I=10^{38}\,{\rm kg\,m^2}$ and place the source at an averaged effective distance of $d=10\,{\rm kpc}$. For a monochromatic GW emission with amplitude $h$ at frequency $f$, we have
\begin{equation}
    C_{12}(f) = \frac{1}{8}T_{\rm FFT} h^2(f).
\end{equation}
Depending on the value of $T_{\rm FFT}$ (and hence the frequency resolution in the CSD), the Doppler effect due to the revolution of the Earth around the Sun and the revolution of the pulsar itself if it is in a binary system can cause the signal to drift in multiple frequency bins.  As a proof-of-concept study, we ignore here the complication due to the Doppler shift as it can be readily corrected for with existing algorithms such as FrequencyHough~\cite{Astone:14} (we would need to include the sky location of the source as search parameters here). Under this assumption, upper limits on the GW strain $h_{\rm UL}$ can then be obtained as (cf. eq. (6) in Ref.~\cite{LVK:22isopulsar}; see also Ref.~\cite{Astone:14})
\begin{equation}
    h_{\rm UL} = 2\sqrt{2{\rm SNR}_{\rm th}\, \frac{|\left(C_{12}\right)_{\rm noise}|}{T_{\rm FFT}}},
    \label{eq:h_UL_CW}
\end{equation}
where ${\rm SNR_{th}}$ is a threshold SNR to claim a detection (note our definition of SNR follows Ref.~\cite{Thrane:11} and is defined in terms of power instead of amplitude as used in Ref.~\cite{Astone:14}). Depending on whether $|\left(C_{12}\right)_{\rm noise}|$ is limited by $\sigma_{C_{12}}$ or $P_{c,{\rm cl}}$ [Eq.~(\ref{eq:C12_noise_CW})], $h_{\rm UL}$ scales with $T_{\rm FFT}$ as $T_{\rm FFT}^{-1/4}$ or $T_{\rm FFT}^{-1/2}$ for given $T_{\rm obs}$. In both cases, longer segment length is preferred. The upper limits are shown in the bottom panel of Fig.~\ref{fig:CW}. Consistent with Ref.~\cite{LVK:22isopulsar}, we have assumed $T_{\rm FFT}=1024\,{\rm s}$. We also adopted a fiducial detection threshold of ${\rm SNR_{th}}=5$. 

Note that when applied to the detection of continuous GW emissions, quantum correlation provides a way to find hot pixels in the $ft$-map, which can then be fed to the FrequencyHough algorithm~\cite{Astone:14} as inputs. It thus complements the existing methods such as using the auto-regressive estimation as proposed in Ref.~\cite{Astone:05}. While quantum correlation does not enhance the fundamental sensitivity, we could potentially benefit from its simplicity in removing the expectation value of the shot noise. Moreover, it naturally handles fluctuations and nonstationarities in the interferometer (as $E_{\rm lo}$ is a common reference field when computing $\la P_a \ra$ and $\la P_b \ra$; see the discussion in \Cref{sec:conclusion_discussion}).
Though as a caveat, to reach the full sensitivity of quantum correlation, it requires the system to be well balanced. We will discuss this point more in \Cref{sec:conclusion_discussion}.

\section{Conclusion and discussion}
\label{sec:conclusion_discussion}
In this work, we explored the possibility of detecting astrophysical GW events using quantum correlation, a technique that has been used by the LIGO instrumentation group to constrain classical noise in the LIGO interferometers~\cite{Martynov:17, Buikema:20}. 
We analyzed in a generic context  the sensitivity of the technique in Sec.~\ref{sec:QC}. 
The main advantage of quantum correlation is that it requires only a single interferometer for the detection (Sec.~\ref{sec:QC_vs_2IFO}), which naturally leads to a higher duty cycle compared to two-interferometer correlation. Moreover, the signals captured by the two PDs in quantum correlation (Fig.~\ref{fig:cartoon}) will share the same antenna response and arrival time, and consequently the detection search can be made more efficient as we do not need to search over extrinsic parameters like the sky location of the event to align the signals (at least for burst signals where the Doppler effect due to Earth's revolution and rotation can be ignored). 
This also allows us to detect highly polarized GW signals with quantum correlation. 
We then considered a few specific examples of using this technique to detect high-frequency GW signals in Sec.~\ref{sec:astro}, including BNS post-merger remnants with both short- ($< 1\,{\rm s}$; Sec.~\ref{sec:short_burst}) and intermediate ($\sim 100\,{\rm s}$; Sec.~\ref{sec:long_burst}) duration, as well as continuous GW emissions from pulsars (especially those at high GW frequencies; Sec.~\ref{sec:cw}). 

% While one can interpret the removal of the quantum shot noise as due to the fact that the shot noise is uncorrelated among the two PDs (Fig.~\ref{fig:cartoon}), 
Conceptually, the quantum correlation technique can be understood in analogy to the correlation between two different interferometers (Sec.~\ref{sec:QC_vs_2IFO}). The quantum shot noise is uncorrelated among the two PDs reading out the GW signal (Fig.~\ref{fig:cartoon}), thereby allowing its removal via cross-correlation. This allows us to adopt results developed for analysing the cross-correlation between different interferometers. Indeed, our definition of the SNR for each pixel [Eq.~(\ref{eq:snr_per_pixel})] follows closely Ref.~\cite{Thrane:11}, and multiple pattern recognition algorithms~\cite{Thrane:13, Thrane:15} can be readily applied to search for the signal in the $ft$-map. 

Meanwhile, Eq.~(\ref{eq:C12_exp}) suggests that we can also view the quantum correlation as follows. By introducing the new field $E_b$, it provides us an estimation of $\la P_a \ra$ (as $\la P_a \ra = \la P_b \ra$), thereby allowing the removal of its expectation value in the (cross-)spectra. This further suggests that if we know $\la P_a \ra$, we may also directly subtract it out and detect the GW signal as excess power in the residual spectra. 
In practice, to do the direct subtraction one would need to take into account the nonstationarity in the interferometer, which could mean extra complications compared to performing quantum correlation.
For example, to directly predict the value of $\la P_a \ra$, one would need to know the instantaneous value of the local-oscillator field $E_{\rm lo}$, whereas the fluctuations in $E_{\rm lo}$ do not affect the quantum correlation because it serves as a common reference when computing $\la P_a \ra$ and $\la P_b \ra$. 
Nevertheless, doing the direct subtraction is an interesting direction to be explored by future studies as it might improve the sensitivity by avoiding the introduction of the uncertainties in $P_b$, which degrades the SNR in Eq.~(\ref{eq:C12_exp}) by $\sqrt{2}$. To tackle the nonstationarities in the interferometer, one could utilize auxiliary channels in LIGO~\cite{Yu:21d, Yu:22a}. Furthermore, with auxiliary channels one could predict not only the expected spectrum of the quantum shot noise $\la P_a \ra$ but also other noise sources across the entire spectra. We plan to explore this possibility in follow-up studies. 

We note that the quantum correlation may pick up instrumental glitches as hot pixels in the CSD. Nonetheless, there are at least two ways to distinguish between terrestrial glitches and astrophysical signals. One is to look for coincidence of hot pixels in multiple detectors. Indeed, this is how we can extract sky location of the source under the quantum correlation technique. On the other hand, if we only find a signal candidate in a single detector, it may still have an astrophysical origin as its disappearance in the other detector(s) might be due to unfavorable antenna response. In this case, we can still veto instrumental glitches utilizing auxiliary channels as routinely done for the LIGO detector characterization~\cite{Essick:13, Essick:20, Davis:22}. 

For quantum correlation, uncertainties in the interferometer calibration~\cite{Sun:20, Sun:21} do not significantly affect the detection. This is because the signal is detected as excess power, which further means the phase of the signal is not used and the amplitude can be measured directly in terms of raw power in the readout PDs to establish the detection statistics. The calibration from power in the PDs to the astrophysical strain will affect mostly the subsequent inference of source parameters such as its distance and ellipticity yet less the detection of the signal itself. %On the other hand, to perform direct subtraction of the expected noise PSD (especially at frequencies where the sensitivity is not limited by the shot noise), the calibration uncertainty of the interferometer is another factor needs to be considered. 

As a proof-of-principle study, we have assumed the final beamsplitter shown in \Cref{fig:cartoon} is an ideal 50-50 beamsplitter, and when squeezed vacuum is used, we have assumed the squeezing level of the $E_b$ field matches exactly the $E_a$ field. In reality, imbalance exists inevitably, which can potentially degrade the performance of the quantum correlation technique. One way to set the tolerance on the imbalance between the beamsplitter's transmissivity and reflectivity and the mismatch of squeezing levels for different squeezers is by requiring [cf. Eq.~(\ref{eq:C12_noise_CW})]
\begin{equation}
    |\la P_a \ra - \la P_b \ra|  \lesssim  {\rm max}\left[ P_{c,{\rm cl}}, \frac{\sigma_{C_{12}}}{\sqrt{N}}\right].
\end{equation}
%This condition is most relevant for the sensitivity to continuous-wave emission as discussed in Sec.~\ref{sec:cw}. 
For aLIGO, the requirement is set by the $P_{c,{\rm cl}}$ term, and satisfying the condition at 1,000 Hz means the difference in $\la P_a \ra$ and $\la P_b \ra$ needs to be $\lesssim 3\%$. 
For CE, $P_{c,{\rm cl}}$ is significantly lower compared to the shot noise and the requirement is set by the number of averages $N$. In this case, the mismatch needs to be $\lesssim 0.4\%$ for $N=10^{5}$.
While failing to meet the requirement above will degrade the sensitivity to continuous-wave emission as discussed in \Cref{sec:cw}, it does not significantly hinder the sensitivity to burst signals (Secs.~\ref{sec:short_burst} and \ref{sec:long_burst}). 
For detecting the burst signals, the requirement is set by 
\begin{equation}
    |\la P_a \ra - \la P_b \ra|  \lesssim  \sigma_{C_{12}}/\sqrt{N_{\rm pix}},
\end{equation}
where $N_{\rm pix}$ is the number of pixels along a detection trajectory. 
Since we typically have $N_{\rm pix}\lesssim 1,000$ for burst signals, we have $\left( \sigma_{C_{12}}/\sqrt{N_{\rm pix}} \right) < \left\la P_{c, {\rm cl}}\right\ra$ especially for future detectors like CE. This leads to a more feasible requirement on the mismatch to be $\lesssim 4\%$.

\begin{acknowledgments}
We thank the helpful comments and feedback from L. Sun, K. Riles and other LVK colleagues. H.Y. is supported by the Sherman Fairchild Foundation. D.M. acknowledges the support of the Institute for Gravitational Wave Astronomy at the University of Birmingham, STFC (Grant No. ST/T006609/1, ST/S000305/1), and EPSRC research councils (Grant No. EP/V048872/1, EP/V008617/1). R.X.A is supported by NSF Grants No. PHY-1764464 and PHY-1912677.  Y.C.\ is supported by the Simons Foundation (Award Number 568762) and by NSF Grants PHY-2011961, PHY-2011968, PHY--1836809.

\end{acknowledgments}

\bibliography{ref}

% \newpage
% \begin{figure*}
%   \centering
%   \includegraphics[width=\textwidth]{supp.pdf}
% \end{figure*}
\end{document}